\documentclass[preprint,aps]{revtex4}
\usepackage{epsfig}

\begin{document}

\title{DYNAMICAL MODEL OF ELECTROWEAK PION PRODUCTION REACTION
}

\author{
T. Sato$^a$ and T.-S. H. Lee$^b$
}

\affiliation{
$^a$
Department of Physics, Osaka University, Toyonaka, Osaka 560-0043,
 Japan\\ 
$^b$
Physics Division, Argonne National Laboratory, Argonne,
Illinois 60439}

\begin{abstract}

The dynamical model of pion electroproduction developed by the
authors has been extended to study the pion weak production reaction.
The axial vector $N-\Delta$  transition form factor $G_A^{N\Delta}$
obtained from the analysis of existing neutrino reaction data and
 the role of dynamical pion cloud on $G_A^{N\Delta}$ are discussed.

\end{abstract}

\maketitle

\section{Introduction}

An important challenge in  hadron physics research is to understand the
hadron structure and reaction dynamics within QCD.
The electroweak form factors of baryon resonance give important
 informations to step forward the research in this direction.
Those informations are important in testing hadron model and
possibly the lattice QCD calculation.
We have developed a dynamical model for investigating 
the pion photoproduction and electroproduction reactions
 in the delta resonance region \cite{sl1,sl2}.
In the dynamical approach\cite{sl1,sl2,sl3,nbl,surya,ky1} we solve the 
scattering equations with the interactions defined from the
effective Lagrangian and/or hadron models. The dynamical
approach is different from the approaches based on the
dispersion relations[7-9] or K-matrix approach[10-13]
 in interpreting the data.
In Refs. 1,2, we have not only extracted the $N\Delta$ transition form factors
 from  the data but also we have provided an interpretation of
the  extracted form factors in terms of hadron models.
It was found in Refs. 1 and 2 that the pion cloud effects give
large contribution to the $N\Delta$ transition form factors. 
The magnetic dipole $N\Delta$ transition form factor $G_M^{N\Delta}(0)$
is enhanced by about 40\% by meson cloud, which gives explanation on the
long standing discrepancy between the prediction of the constituent
quark model and empirical amplitude analyses. Furthermore,
the long range pion cloud gives soft component of the
$G_M^{N\Delta}(Q^2)$ and very pronounced enhancement of quadrupole
transition form factors $G_E^{N\Delta}(Q^2), G_C^{N\Delta}(Q^2)$
at low $Q^2$. 

The axial vector response of the hadron can be studied by the weak
processes. The neutrino-induced pion production reaction
$\nu + N \rightarrow l + \pi + N$ can be used to extract such informations.
The previous investigations of the weak pion production reaction[14-22]
 have been done  by using dispersion relation approach or
K-matrix approach. 
We report on our recent progress on the neutrino-induced
pion production reaction in the delta resonance region\cite{sl3}.
The purpose of this work is to develop a dynamical model for
neutrino-induced pion production reaction to extract axial vector
$N\Delta$ form factors by extending our model on pion electroproduction.
In particular we investigate possible role of the dynamical pion
cloud in solving the problem that the $N\Delta$ axial vector form factor
extracted from  neutrino reaction\cite{hhm,lmz} was found to be
 about 30\% larger than the quark model predictions.
In section 2, we will briefly describe our dynamical model.
The results of neutrino-induced pion production and the axial vector
$N\Delta$ form factor is given in section 3.

\section{Dynamical approach}
We start from the Hamiltonian of mesons and baryons fields with the 
interaction Hamiltonian $H_I= \sum_{B',M',B}\Gamma^0_{B'M',B}$, 
which describes absorption and emission of meson $M$ from baryon $B$ as
\begin{eqnarray}
H & = & H_0 + \Gamma_{B'M',M} + (\mbox{h.c.}).
\end{eqnarray}
Similarly the weak hadron current $J^\mu$ consists of
mesons and  baryons weak currents. To obtain a manageable reaction 
theory to describe neutrino-induced pion production reaction, we apply unitary
transformation method\cite{sl1,ksh} up to the second order of the
interaction Hamiltonian.
The idea is that we eliminate 'virtual' interaction $B \rightarrow B'M'$
for $m_B < m_M' + m_B'$ from the Hamiltonian in Eq. (1)
and absorb their effects into many-body potentials.
As a result $N$ and $\pi N$ states decouple with each other up to the
order of our approximation and the effective Hamiltonian consists of 
the interactions of the resonance decay/production and the many-body 
potentials.
\begin{eqnarray}
H_{eff} = H_0 + v_{\pi N} + \Gamma_{\Delta, \pi N} + \mbox{h.c.}.
\end{eqnarray}
\begin{figure}[h]
\centerline{\epsfig{file=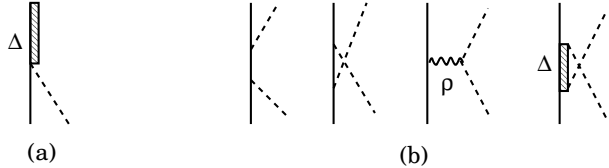,width=8cm}}
\vspace*{8pt}
\caption{Graphical representation of the interactions
(a) $\Gamma_{\Delta,\pi N}$ and (b) $v_{\pi N}$.}
\end{figure}
\begin{figure}[h]
\vspace*{8pt}
\centerline{\epsfig{file=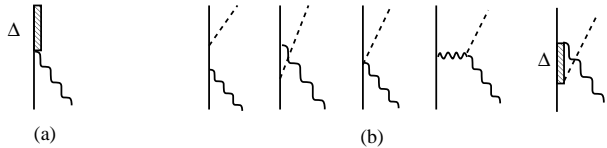,width=8cm}}
\caption{Graphical representation of the effective current
(a) $J^\mu_{\Delta N}$ and (b) $J^\mu_{\pi N}$.}
\end{figure}
Here $v_{\pi N}$ is  non-resonant $\pi N$ potential and
 $\Gamma_{\Delta, \pi N}$ describes  decay of delta into $\pi N$.
The effective hadron current $J^\mu_{eff}$ is obtained by applying the
same unitary trasformation used for $H_{eff}$ on $J^\mu$.
 $J^\mu_{eff}$ consists of the non-resonant pion production current
$J^\mu_{\pi N}$ and $N\Delta$ transition current $J^\mu_{\Delta N}$:
\begin{eqnarray}
J^\mu_{eff} = J^\mu_{\Delta N} + J^\mu_{\pi N}.
\end{eqnarray}
From the effective Hamiltonian and current,
the T-matrix of the pion production reaction can be
easily obtained by solving the coupled channel
Lippman-Schwinger equation within the $\pi N \oplus \Delta$ Fock space.
The resulting matrix element of the current  $\bar{J}^\mu_{\pi N}$,
which includes final state interaction and
 satisfies  the Watson theorem, is given as
\begin{eqnarray}
\bar{J}^\mu_{\pi N} = \bar{J}^\mu_{\pi N}(\mbox{non-res})
 + \frac{\bar{\Gamma}_{\Delta,\pi N}
\bar{J}^\mu_{\Delta N}}{W - m_\Delta - \Sigma}.
\end{eqnarray}
The non-resonant current $\bar{J}^\mu_{\pi N}(\mbox{non-res})$
is calculated only from non-resonant interactions $v_{\pi N}$ and 
 $J^\mu_{\pi N}$. 
The second term in Eq. (4) is resonant amplitude
with the dressed  $J^\mu + N \rightarrow \Delta$ vertex
 given as
\begin{eqnarray}
\bar{J}^\mu_{\Delta N}(W,q) & = & J^\mu_{\Delta N} +
 \int dk k^2 
\frac{\bar{\Gamma}_{\Delta, \pi N}(W,k)J^\mu_{\pi N}(k,q)}
     {W - E_N(k) - E_\pi(k) + i \epsilon}.
\end{eqnarray}
An important feature of the dynamical model is that the
bare vertex  $J^\mu_{\Delta N}$, which
 may be compared with the prediction of the hadron model,
is  modified by the off-shell non-resonant interaction $J^\mu_{\pi N}$
to give the dressed vertex $\bar{J}^\mu_{\Delta N}$.

\section{Weak pion production reaction}

We apply the method described in the previous
section to  $\nu_\mu + N \rightarrow \mu + \pi + N$ reaction.
The low energy effective Lagrangian of the electroweak
standard model is given as
\begin{eqnarray}
 H_{W} & = & \frac{G_F \cos_c}{\sqrt{2}}(V^\mu - A^\mu)l_\mu .
\end{eqnarray}
Here $V^\mu$ and $A^\mu$ are charged currents of hadron and
$l^\mu$ is lepton current. The vector current($V^\mu$) is obtained from
the electromagnetic current in Refs. 1,2 by assuming
CVC and iso-spin rotation. The axial vector hadron current is 
obtained from  the chiral Lagrangian. 
Following the procedure described in the previous section,
the pion production currents $J^\mu_{eff}$ are calculated
shown in the schematic diagrams of Fig. 2.
We took the constituent quark model relation of 
the  coupling constant
of the  axial $N\Delta$ current $A_{\Delta N}^\mu$  is obtained from the
quark model relation $G_A^* = \sqrt{72/25} g_A$.
The $Q^2$ dependence of the axial form factor is assumed as
\begin{eqnarray}
G_A^*(Q^2) = G_A^(0)\frac{1}{(1 + Q^2/m_A^2)^2}R_{SL}(Q^2),
\end{eqnarray}
where the dipole form factor
 is axial vector form factor of nucleon  with $m_A= 1.02GeV$\cite{meissner}
and phenomenological form factor $R_{SL}(Q^2)= (1 + a Q^2)exp(- b Q^2)$
with $a=0.154 GeV^{-2}$
  and $b=1.66 GeV^{-2}$ is obtained by fitting pion electroproduction 
reaction cross section at $Q^2=2.8$ and $4(GeV/c)^2$. 
All the other coupling constants and the form factors used in
this work are the same value as our model of the pion electroproduction.
Therefore, there is no adjustable parameters  in this calculation.

We first compare the total cross section of $\nu_\mu + p \rightarrow \mu^-
+ \pi^+ + p$ reaction with the data\cite{data79} in Fig. 3. 
Our result shown in solid curve agrees reasonably
well with the data. One of the main feature of the dynamical approach is
the bare form factor of $N\Delta$ transition is renormalized by the pion
rescattering effects. The importance of this effect is seen
in Fig. 3. 
The full result in solid line is reduced 
into dashed line 
when we turn off the dynamical pion cloud effect.
The contribution of the pure non-resonant contribution 
is shown in dot-dashed line. We next compare the $Q^2$ dependence of the
calculated differential cross section with the data \cite{data90} in Fig. 4.
Our model reproduce very well the data. In the low $Q^2$ region, the
main contribution is the the axial vector current shown in the dashed
line. The BNL data were used in the most recent attempt to extract
$N\Delta$ axial vector form factor.

Finaly we study the effect of dynamical pion cloud effect on the axial
vector form factor. The empirical $N\Delta$ from factor extracted from
the data can be compared with our dressed form factor
$\bar{J}^\mu_{\Delta N}$ shown in the solid line in Fig. 5, while the
contribution of the bare form factor $J^\mu_{\Delta N}$ is shown in the
dotted line. We see sizable contribution of the dynamical pion cloud as
the case of  the electromagnetic form factors. 
The differences between solid and
 dotted lines explain the observation that 
the quark model prediction of axial $N\Delta$ from
 factor is about 30\% smaller than the empirically extracted form factor.

\begin{figure}
\centerline{\epsfig{file=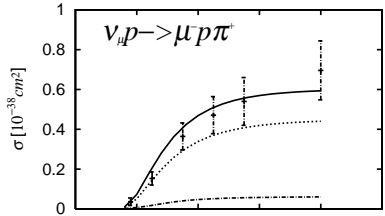,width=5cm}}
\vspace*{8pt}
\caption{Total cross section of $p + \nu_\mu \rightarrow \mu^- + \pi^+ + p$
    reaction.}
\end{figure}
\begin{figure}
\centerline{\epsfig{file=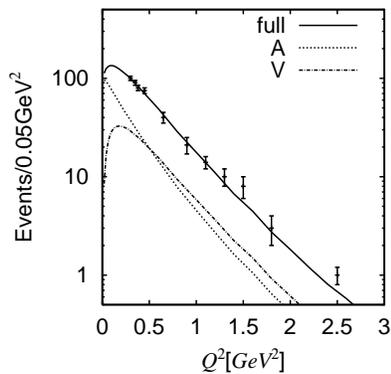,width=5cm}}
\vspace*{8pt}
\caption{Differential cross section $d\sigma/dQ^2$ of
 $p + \nu_\mu \rightarrow \mu^- + \pi^+ + p$ reaction}
\end{figure}

\begin{figure}
\centerline{\epsfig{file=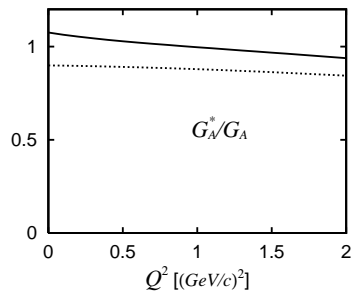,width=5cm}}
\vspace*{8pt}
\caption{The $N\Delta$ axial vector form factor.}
\end{figure}

\section{Summary}

In summary, the dynamical model developed in \cite{sl1,sl2,sl3} can
describe extensive data of pion photoproduction, electroproduction
and the neutrino-induced reactions. The predicted $\nu_\mu + N \rightarrow
\pi + N + \mu$ cross sections are in good agreement  with the existing
data. The renormalized axial vector $N\Delta$ form factor contains 
large dynamical pion cloud effects and these effects are crucial in
getting agreement with the data. We conclude that the $N\Delta$ transition
axial vector form factor predicted from the quark model is consistent
with the existing data in the delta region. However more extensive and
precise data of neutrino-induced pion production reactions are needed to 
further test our model and to pin down the $Q^2$ dependence of the
form factors. The efforts to extend  the current dynamical model
beyond the delta resonance region\cite{sl5} and to improve the effective
Hamiltonian including leading loop corrections\cite{sl4} are in progress.

\section*{Acknowledgments}

This work was supported by the 
Japan Society for the Promotion of Science, 
Grant-in-Aid for Scientific Research (C) 15540275 and
 the U.S. Department of Energy, Office of
Nuclear Physics Division, under contract no. W-31-109-ENG-38.


\begin{thebibliography}{0}

\bibitem{sl1}
    T.~Sato and T.-S.~H. Lee, Phys. Rev. C {\bf 54}, 2660 (1996).
\bibitem{sl2}
    T.~Sato and T.-S. H. Lee, Phys. Rev. C {\bf 63}, 055201 (2001).

\bibitem{sl3}
    T.~Sato, D.~Uno  and T.-S.~H. Lee, Phys. Rev. C {\bf 67}, 065201 (2003).

\bibitem{nbl}
S. Nozawa, B. Blankleider, and T.-S. H. Lee, Nucl. Phys. {\bf A513},
459 (1990).

\bibitem{surya}
Y. Surya and F. Gross, Phys. Rev. C {\bf 53}, 2422 (1996).

\bibitem{ky1}  S. S. Kamalov and S. N. Yang,
                Phys. Rev. Lett. {\bf 83}, 4494 (1999). 



\bibitem{cgln}  G.F. Chew, M.L. Goldberger, F.E. Low and Y. Nambu,
Phys. Rev. {\bf 106}, 1345 (1957).

\bibitem{hdt} O. Hanstein, D. Drechsel and L. Tiator,
                  Nucl. Phys. {\bf A632}, 561 (1998).

\bibitem{a98} I. G. Aznauryan,
                    Phys. Rev. D {\bf 57}, 2727 (1998).

\bibitem{muko90} R. M. Davidson and N. C. Mukhopadhyay,
                 Phys. Rev. D {\bf 42}, 20 (1990);\\
                 R. M. Davidson, N. C. Mukhopadhyay and R. S. Wittman,
                 Phys. Rev. D {\bf 43}, 71 (1991).


\bibitem{muko99}  R. M. Davidson et al.,
                 Phys. Rev. C {\bf 59}, 1059 (1999).

\bibitem{maid} D. Drechsel, O. Hanstein, S.S. Kamalov and L. Tiator,
               Nucl. Phys. {\bf A645}, 145 (1999).

\bibitem{ky2}  S. S. Kamalov, S. N. Yang, D. Drechsel,
   O. Hanstein, and L. Tiator, Phys. Rev. C {\bf 64}, 032201(R) (2002).


\bibitem{salin}
P. Salin, Nuovo Cimento A {\bf 48}, 506 (1967).

\bibitem{adler}
S. L. Adler, Ann. Phys. {\bf 50}, 189 (1968); Phys. Rev. D {\bf 12},
2644 (1975).

\bibitem{bij}
J. Bijtebier, Nucl. Phys. {\bf B21}, 158 (1970).

\bibitem{zuker}
P.A. Zuker,  Phys. Rev. D {\bf 4}, 3350 (1971).

\bibitem{schr73}
P. A. Schreiner and F. Von Hippel, Nucl. Phys. {\bf B58}, 333 (1973).

\bibitem{rein}
D. Rein and L. M. Sehgal, Ann. Phys. {\bf }, 79 (1980).

\bibitem{sakuda}
E. A. Paschos, M. Sakuda, I. Schienbein and J. Y. Yu, hep-ph/0408185 (2004).

\bibitem{hhm}
T.R. Hemmert, B.R. Holstein, and N.C. Mukhopadhyay, Phys. Rev. D
{\bf 51}, 158 (1995).

\bibitem{lmz}
J. Liu, N.C. Mukhopadhyay, and L. Zhang, Phys. Rev. C {\bf 52}, 1630 (1995).

\bibitem{ksh}  M. Kobayashi, T. Sato and H. Ohtsubo,
               Prog. Theor. Phys. {\bf 98}, 927 (1997).


\bibitem{meissner}
V. Bernard et al., J. Phys. {\bf G28}, R1 (2002).

\bibitem{data90}
T. Kitagaki et al, Phys. Rev. D {\bf 42}, 1331 (1990).


\bibitem{data79}
S.J. Barish et al., Phys. Rev. D {\bf 19}, 2521 (1979).


\bibitem{sl5}
    A.~Matsuyama, T.-S.~H. Lee, and T.~Sato, nucl-th/0406050 (2004).


\bibitem{sl4}
    T.~Sato,  T.-S.~H. Lee, and T.~Nakamura, nucl-th/0411013 (2004).

\end{thebibliography}
\end{document}